\documentclass[aps,pra,preprint,superscriptaddress]{revtex4-1}
\usepackage{graphicx}
\usepackage{amsfonts}
\usepackage{amsmath}
\usepackage{color}
\usepackage{bm}
\usepackage{amssymb}

\usepackage{multirow}
\usepackage{natbib}

\newcommand{\beq}{\begin{equation}}
\newcommand{\eeq}{\end{equation}}

\def\jcp#1#2#3{{J.~Chem.~Phys.}~{\bf #1},\ #2\ (#3)}

\def\prl#1#2#3{{Phys.~Rev.~Lett.}~{\bf #1},\ #2\ (#3)}

\def\colvecnext#1{
        #1
        \global\advance\colveccount-1
        \ifnum\colveccount>0
                \\
                \expandafter\colvecnext
        \else
                \end{pmatrix}
        \fi
}

\begin{document}

\title{Cold collisions of heavy $^2\Sigma$ molecules with alkali-metal atoms in a magnetic field: {\it Ab initio} analysis and prospects for sympathetic cooling of SrOH$(^2\Sigma)$ by Li($^2$S) }
%\title{Molecular collisions and reactive scattering in external fields: Field-induced couplings at short range}
\author{Masato Morita}
\affiliation{Department of Physics, University of Nevada, Reno, NV, 89557, USA}
\author{Jacek K{\l}os}
\affiliation{Department of Chemistry and Biochemistry, University of Maryland, College Park, MD, 89557, USA}
\author{Alexei A. Buchachenko}
\affiliation{Skolkovo Institute of Science and Technology, Skolkovo Innovation Center, Building 3, Moscow 143026, Russia}
\affiliation{Department of Chemistry, M. V. Lomonosov Moscow State University, Moscow 119991, Russia}
%\affiliation{Institute of Problems of Chemical Physics RAS, Chernogolovka, Moscow Region 142432, Russia}
\author{Timur V. Tscherbul}
\affiliation{Department of Physics, University of Nevada, Reno, NV, 89557, USA}

\date{\today}

\begin{abstract}
 We use accurate {\it ab initio} and quantum scattering calculations to explore the prospects for sympathetic cooling of the heavy molecular radical SrOH($^2\Sigma$) by ultracold Li atoms in a magnetic trap. A two-dimensional potential energy surface (PES) for the  triplet electronic state of Li-SrOH is calculated {\it ab initio} using the partially spin-restricted coupled cluster  method with single, double and perturbative triple excitations and a large correlation-consistent basis set. The highly anisotropic PES has a deep global minimum  in the skewed Li-HOSr geometry with $D_e=4932$ cm$^{-1}$ and saddle points in collinear configurations.
 Our quantum scattering calculations predict low spin relaxation  rates in fully spin-polarized Li~+~SrOH collisions with the ratios of elastic to inelastic collision rates well in excess of 100 over a wide range of magnetic fields (1-1000 G) and collision energies (10$^{-5}-0.1$~K) suggesting favorable prospects for sympathetic cooling of SrOH molecules with spin-polarized Li atoms in a magnetic trap. We find that spin relaxation in Li~+~SrOH collisions occurs via a direct  mechanism  mediated by the magnetic dipole-dipole interaction between the electron spins of Li and SrOH, and that the indirect (spin-rotation) mechanism  is strongly suppressed. The upper limit to the Li~+~SrOH reaction rate coefficient calculated for the singlet PES using adiabatic capture theory is found to decrease from $4\times 10^{-10}$~cm$^3$/s to a limiting value of $3.5\times 10^{-10}$ cm$^3$/s with decreasing temperature from 0.1 K to 1 $\mu$K. 
  \end{abstract}

\maketitle

\section{Introduction}

Ultracold molecular gases offer a wide range of  research opportunities, extending from quantum simulation of many-body systems with long-range dipolar interactions \cite{njp09,ChemRev12,QuantumSimulation} to external field control of  chemical reaction dynamics \cite{RomanPCCP08,JunARPC14,ChemRev12}, precision measurement of molecular energy levels to uncover new physics beyond the Standard Model \cite{AndreiMatrix,ThO,JunOH,Ubachs}, and quantum information processing with molecular arrays in optical lattices \cite{DeMille06}. At present, coherent association of ultracold alkali-metal atoms remains the only experimental technique to produce ultracold  molecular gases of  KRb and NaK  \cite{KRb1,KRb2,NaK1}. Recent advances in  laser cooling and magneto-optical trapping \cite{DeMille1,DeMille2,DeMille3,YO1,YO2},  single-photon cooling \cite{Ivan14}, Sisyphus laser cooling  \cite{Ivan3,Rempe1}, and optoelectrical cooling  \cite{Rempe1,Rempe2} made it possible to control and confine  molecular species such as SrF, CaF, SrOH, YO, CH$_3$F, and H$_2$CO in electrostatic and magnetic traps at temperatures as low as a fraction of a milliKelvin \cite{DeMille1,DeMille2,DeMille3,YO1,YO2,Ivan14,Rempe1,Rempe2}. %Using these direct cooling techniques, several experimental groups have brought  molecular species SrF \cite{DeMille1,DeMille2}, CaF \cite{Ivan14}, YO \cite{YO1,YO2}, SrOH \cite{IvanSrOH1,IvanSrOH2,IvanSrOH3}, and H$_2$CO \cite{Rempe} to sub-mK temperatures.
Due to the intrinsic limitations of optical cooling, it is necessary to employ second-stage cooling techniques to further reduce the temperature of a trapped molecular gas to  $<$0.1~mK \cite{njp09,Ivan14}.

One such  technique---sympathetic cooling---relies on elastic atom-molecule collisions to transfer energy and momentum from a cold molecular gas to an ultracold reservoir of neutral atoms.  While elastic collisions drive momentum-transfer and thermalization, inelastic collisions release the internal energy of the molecules,  leading to  heating and ultimately trap loss. In order to remain trapped in the inhomogeneous magnetic field of a conservative magnetic trap, open-shell molecules must reside in the low-field-seeking Zeeman states, which contain an excess of internal Zeeman energy. This energy can be released in collisions with buffer-gas atoms in a process  known as collision-induced spin relaxation \cite{njp09,RomanPCCP08}.  For sympathetic cooling experiments, it is desirable to keep the molecules in the trap for as long as possible; hence, the rate of collision-induced spin relaxation should be small  compared to the elastic collision rate. To allow for efficient thermalization of trapped molecules on the experimental timescale, the ratio of elastic to inelastic collision rates  should exceed 100 \cite{njp09}.

%in conservative (e.g. magnetic)  traps  it is important e elastic, collision-induced spin relaxation s technique relies on elastic collisions that transfer momentum from molecules to a pre-cooled gas of ultracold atoms.  to thermalize a molecular gas with a much colder ultracold  . The collisional properties are poorly understood.... This paragraph - 

%Recent advances in laser cooling and magneto-optical trapping of molecular radicals EXPERIMENTAL WORK
%In addition, 

Several groups have  explored the possibility of using ultracold alkali-metal atoms to sympathetically cool  paramagnetic  molecules such as OH, NH, CaH, and CaF in a magnetic trap using accurate {\it ab initio} and quantum scattering calculations. In particular, Lara {\it et~al.} showed that inelastic relaxation in cold Rb~+~OH collisions occurs at a high rate, thereby precluding sympathetic cooling of magnetically trapped OH by Rb atoms \cite{Lara1}. The spin relaxation cross sections for collisions of polar radicals NH and OH with spin-polarized N and H atoms were found to be small owing to the weak anisotropy of the high-spin NH-N, NH-H, OH-H interactions, making atomic nitrogen and hydrogen promising coolant atoms \cite{N-NH,Piotr-N-NH,Piotr-N-OH,H-OH}. The same conclusion was reached for ground-state Mg atoms colliding with NH$(^3\Sigma)$  molecules \cite{Mg-NH}. We showed that despite the strong angular anisotropy of the interactions between $^2\Sigma$ molecular radicals and alkali-metal atoms, the inelastic cross sections for interspecies collisions are  strongly suppressed due to the weakness of the spin-rotation interaction in $^2\Sigma$ molecules \cite{pra11}. Small polyatomic molecular radicals such as methylene (CH$_2$), methyl (CH$_3$), and amidogen (NH$_2$)  were found to have small spin relaxation cross sections with $S$-state atoms, and hence suggested  as promising candidates for sympathetic cooling experiments in a magnetic trap \cite{prl11,jcp12}. 

%nelastic cross sections calculated for  restricted basis sets, so accurate predictions could me made  that did not guarantor numerical convergence.

The vast majority of atom-molecule combinations  proposed for sympathetic cooling experiments  included light hydrogen-containing molecules such as NH, OH, and CaH. These molecules have large rotational level spacings  and low densities of rovibrational states, facilitating accurate quantum scattering calculations \cite{jcp12a,pra11}. In contrast, the heavy molecular radicals produced and studied in recent experiments (CaF, SrF, YO, and SrOH) have small rotational constants and dense spectra of rovibrational states. 
While the possibility of co-trapping and sympathetic cooling of $^2\Sigma$ molecular radicals with ultracold alkali-metal atoms has been suggested \cite{pra11,Ivan14,Ivan3}, numerically exact quantum scattering calculations of their collisional properties are challenging \cite{pra11} due to the strongly anisotropic atom-molecule interactions, which couple a large number of rovibrational states and field-induced mixing between different total angular momenta (see Ref. \cite{jpb16} for a detailed discussion). As a result, it remains unclear whether heavy molecular radicals trapped in recent experiments \cite{DeMille1,DeMille2,DeMille3,YO1,YO2,Ivan14}  have small enough inelastic collision rates with ultracold alkali-metal atoms  to allow for efficient sympathetic cooling  in a magnetic trap.

   Cooling and trapping polyatomic molecular radicals is expected to provide new insights into many-mode vibrational dynamics, photochemistry, and chemical reactivity at ultralow temperatures \cite{prl11,Ivan1,Ivan2,Ivan3,IvanCPC}.   Recently, Kozyryev {\it et al.}  used buffer-gas cooling to prepare a cold sample of the strontium monohydroxide radical [SrOH ($X^2\Sigma$)]  in the ground and first excited vibrational states and to observe vibrational energy transfer between the states induced by  collisions with He atoms at 2~K  \cite{Ivan1}. The highly diagonal array of Franck-Condon factors between the ground $\tilde{X}^2\Sigma^+$ and the first excited $\tilde{A}\Pi_{1/2}$ electronic states of SrOH enables  efficient photon cycling, making SrOH an attractive candidate for molecular laser cooling and trapping. In a series of recent experiments,  Kozyryev {\it et al.} observed the radiation pressure force and demonstrated  Sisyphus laser cooling of SrOH to below 1 mK in one dimension   \cite{Ivan2,Ivan3}.

Here, we use accurate {\it ab initio} and quantum scattering calculations to explore the possibility of sympathetic cooling of SrOH$(^2\Sigma)$ with ultracold Li($^2$S) atoms in a magnetic trap. To this end, we develop an {\it ab initio}  potential energy surface (PES) for the triplet electronic state of Li-SrOH (Sec. IIA) and employ it in multichannel quantum scattering calculations using a computationally efficient total angular momentum representation for molecular collisions in magnetic  fields \cite{jcp10} (Sec. IIB). In Sec. IIIA we show that inelastic spin relaxation of  spin-polarized SrOH molecules in collisions with spin-polarized Li atoms occurs 100-1000 times slower than elastic collisions over a wide range of collision energies and magnetic fields, suggesting good prospects of  sympathetic cooling of SrOH molecules with ultracold Li atoms in a magnetic trap. We find broad resonance features in the magnetic field dependence of atom-molecule scattering cross sections and show that spin relaxation in cold Li~+~SrOH collisions occurs predominantly due to the magnetic dipole-dipole interaction (direct mechanism) rather than via the intramolecular spin-rotation interaction combined with the anisotropy of the interaction potential (indirect mechanism).  In Sec. IIIB we use  adiabatic capture theory to estimate the upper limit to the rate of the Li + SrOH $\to$ LiOH~+~Sr chemical reaction.    The paper concludes in Sec. IV with a  summary of main results and a brief outline of future research directions. Atomic units are used throughout the rest of the paper unless otherwise stated.

\section{Theory and computational methodology}

\subsection{Ab initio calculations}

The SrOH radical is a linear molecule in its ground electronic state of ${}^2\Sigma$ symmetry \cite{SrOHspectroscopy1}. The interaction with a ground-state Li($^2$S) atom gives rise to two adiabatic PESs  of singlet and triplet spin multiplicities. 
The triplet-singlet couplings have been shown to be negligible  in a closely related Li-CaH system   \cite{Warehime:2015} and since our interest here is in collisions of fully spin-polarized Li and SrOH which occur on the triplet PES, we make the common  assumption of neglecting the difference between the singlet and triplet PESs \cite{N-NH,Piotr-N-NH,Piotr-N-OH, H-OH,pra11}.
   This has the added advantage that single-reference electronic structure methods can be used to describe  the triplet   state of the Li-SrOH collision complex. To compute the triplet  PES, we thus employ the partially spin-restricted coupled cluster method~\cite{Knowles:1993} with single, double and perturbative triple excitations (RCCSD(T)) with the reference wavefunction taken from the restricted Hartree-Fock (RHF) approach. The RHF wavefunction was calculated using  pseudocanonical orbitals from multi-reference self-consistent field~\cite{Werner:1985,Knowles:1985} (MCSCF) calculations with valence active space as a starting point. 

The geometry of the complex is described by the Jacobi coordinates $R$ -- the distance between Li and the center of mass of SrOH, $r$ -- the SrOH bond length, and $\theta$ -- the angle between the Jacobi vectors $\bm{R}$ and $\bm{r}$. The origin of the coordinate system is taken at the center of mass of  SrOH. The geometry of SrOH is kept linear and fixed throughout  the calculations. The position of the center of mass was calculated using the exact mass of the most abundant isotope $^{88}$Sr. The Jacobi angle $\theta$ describes the angular dependence of the PES and the $\theta=0^{\circ}$ geometry describes the Li--H-O-Sr collinear arrangement. The linear Sr--O--H geometry  is described by the bond lengths $r(\mathrm{SrO})=2.1110\;$\AA\space and $r(\mathrm{OH})=0.9225\;$\AA\space as verified by the geometry optimization  at the RCCSD(T) level. The normal modes of  SrOH are determined from the vibrational frequency RCCSD(T) calculations with the same basis set as the PES calculations (excluding the bond functions), which show the doubly degenerate SrOH bending mode at 386 cm$^{-1}$, the Sr-O stretching mode at 534~cm$^{-1}$ and the OH stretching vibration at 3919 cm$^{-1}$.

\begin{figure}[ht]
\label{figure1}
\begin{center}
\includegraphics[height=0.65\textheight,keepaspectratio]{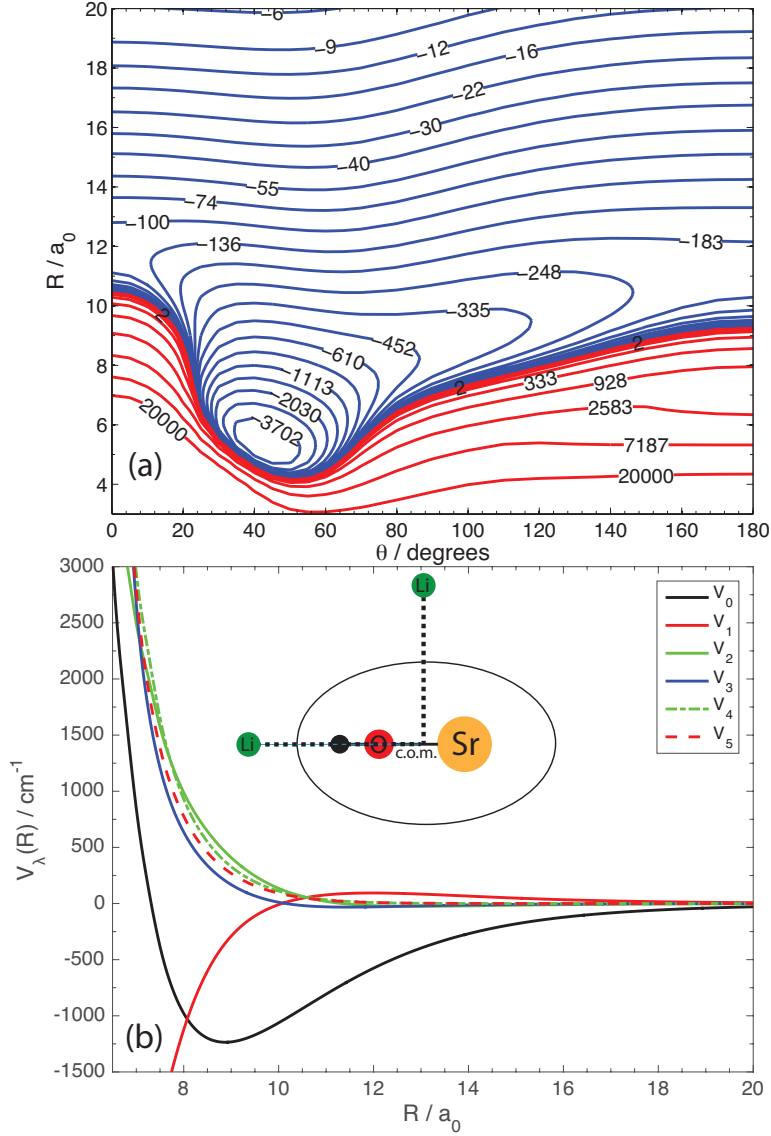}
\end{center}
\caption{\label{pes_fig}(a) A contour plot of the {\it ab initio} potential energy surface for Li-SrOH in its triplet electronic state (in units of cm$^{-1}$). The $\theta=0^{\circ}$ geometry corresponds to the collinear Li--H-O-Sr arrangement. (b) The radial dependence of  the first few Legendre expansion coefficients $V_\lambda(R)$. The insert  shows the ellipsoid along which the bond functions are placed. The center of the ellipsoid is located at the center of mass of SrOH, and its horizontal and vertical axes are given by $r_b=R_\text{Li-H}/2+r_\text{H-X}$ and $r_a=R/2$, where $r_\text{H-X}$ is the distance from H to the center of mass of  SrOH.}
\end{figure}

For the Sr atom, we use a pseudopotential-based augmented correlation-consistent quintuple-zeta basis (aug-cc-PV5Z-PP) of Peterson and coworkers~\cite{Sr_av5zpp_basis} with Stuttgart/Cologne effective core potential (ECP) (ECP28MDF)~\cite{Sr_ECP28MDF}. The remaining  Li, O, and H atoms are described by core-valence Dunning's  aug-cc-pCVTZ basis  functions~\cite{Dunning:1989}. The basis set used in the calculations of the Li-SrOH complex is augmented with a  set of $3s3p2d2f1g$ bond-functions placed on an ellipsoid as shown in the inset of Fig.~\ref{pes_fig} \cite{Prosmiti}. The bond functions were represented using the following exponents: $sp=(0.94,0.34,0.12)$, $df=(0.64, 0.23)$ and $g=(0.35)$. The ellipsoidal placement of the mid-bond functions avoids accidental overlap with the atomic basis  functions of SrOH  at small $R$.

We calculate the triplet PES on a two-dimensional grid in $R$ and $\theta$ within a supermolecular approach and correct for the basis set superposition error using the counterpoise correction procedure of Boys and Bernardi~\cite{Boys:1970}:
\begin{equation}
V(R,\theta)=E_{\mathrm{Li-SrOH}}(R,\theta)-E_{\mathrm{Li-ghost}}(R,\theta)-E_{\mathrm{SrOH-ghost}}(R,\theta).
\end{equation}   
The ghost in the above equation denotes the presence of dimer-centered basis  functions during the calculations of monomer energies. The $R$ Jacobi coordinate is represented by the radial grid of 105 points spanning distances from $R=2.75$ $a_0$ to $R=40$ $a_0$ with a variable step from 0.05 $a_0$ in the medium-range to 0.5-2.0 $a_0$ in the long-range. The angular  variable $\theta$ was represented on a grid of 26 points, with a step of 5 degrees from 0 to 70 degrees and with the step of 10 degrees in the remaining interval to 180 degrees. This gives around 2700 points representing the triplet Li-SrOH PES for a fixed SrOH geometry. All electronic structure calculations have been performed with the MOLPRO suite of programs~\cite{Molpro2012,Molpro-WIREs}.     

The calculated PES data points are expanded in Legendre polynomials 
\begin{equation}
\label{pes_expansion}
V(R,\theta) = \sum_{\lambda=0}^{20}V_{\lambda}(R)P_\lambda(\cos\theta).
\end{equation}
Figure~\ref{pes_fig}(a)  shows a contour plot of the triplet Li-SrOH PES. The potential is extremely anisotropic, varying from strongly attractive (thousands of cm$^{-1}$) in the region of the global minimum to weakly attractive (-100-200 cm$^{-1}$)  near the collinear saddle points at $\theta=0^\circ$ or $\theta=180^\circ$ and $R\approx12$ $a_0$.
 The high anisotropy is also manifested in the large magnitude of the first few anisotropic Legendre moments $V_{\lambda}(R)$ shown in Fig.~\ref{pes_fig}(b) at medium and short $R$. Higher-order Legendre terms become progressively less important at larger atom-molecule separations.  
 The long-range fit is performed using the analytical formula $V_\text{LR}(R,\theta)=-\sum_{n,l}\frac{C_{nl}}{R^n}P_l(\cos\theta)$ including the dispersion coefficients from $C_{60}$ to $C_{84}$. The isotropic van der Waals dispersion coefficient of the triplet PES is estimated from the long-range fit to be $C_{60}=1.7\times 10^9$ cm$^{-1}a_0^6$.  The long-range fit is smoothly joined with the expansion fit [Eq.~(\ref{pes_expansion})] by the hyperbolic tangent switching function. The radial $V_{\lambda}$ coefficients are fit using the Reproducing Kernel Hilbert Space (RKHS) interpolation method with a one-dimensional radial kernel with $n=2$ and $m=5$ \cite{Ho:1996,Hollebeek:1999}. A Fortran routine  for the Li-SrOH PES is available in the  Supplemental Material \cite{SI}. 

The global minimum of the triplet Li-SrOH PES is located at $R_e=5.289$ $a_0$ and $\theta_e=43.19^{\circ}$ with a well depth of 4931.94 cm$^{-1}$. As shown in Fig.~\ref{pes_fig} the global minimum of the Li-SrOH complex corresponds to a skewed Li--HOSr geometry with the Li--H distance of 3.818 $a_0$ and the Sr-O-H--Li  angle of $\approx71.5^{\circ}$. 

% \textcolor{red}{JACEK, PLEASE DESCRIBE YOUR AB INITIO CALCULATIONS and PES GEOMETRY (FIG. 1)} %in the linear $C_{\infty h}$ geometry of the Li-SrOH complex

\subsection{Quantum scattering calculations}

In order to solve the quantum scattering problem for Li-SrOH, we numerically integrate the close-coupling (CC) equations in the body-fixed (BF) coordinate frame \cite{jcp10,pra11}. Motivated by the need to reduce the computational cost of quantum scattering calculations, we assume  that SrOH remains frozen at its ground-state equilibrium configuration, thereby invoking the rigid-rotor approximation \cite{pra11}. The energy gap between the ground and the lowest excited vibrational states of SrOH (386 cm$^{-1}$) is small compared to the Li-SrOH potential strength (4931.9 cm$^{-1}$), which may lead to a temporary excitation of the vibrational modes, giving rise to a resonance structure in the energy and field dependence of scattering cross sections. At collision energies far detuned from the resonances, the coupling between the different vibrational modes of SrOH induced by the interaction with the incident Li atom is small and the rigid-rotor approximation is expected to hold. We therefore expect that our calculations provide a sufficiently accurate description of cold Li~+~SrOH background scattering.

The effective Hamiltonian for low-energy collisions between an atom A ($^2$S) and a diatomic molecule B ($^2\Sigma$) in the presence of an external magnetic field may be written \cite{pra11,jcp10} 
\begin{equation}
\hat{\mathcal{H}} = - \frac{1}{2\mu} R^{-1} \frac{d^2}{dR^2} R
                     + \frac{(\hat{J}-\hat{N}-\hat{S}_{\mathrm{A}}-\hat{S}_{\mathrm{B}})^2}{2\mu R^2} \allowbreak
                     + \hat{\mathcal{H}}_{\mathrm{A}}
                     + \hat{\mathcal{H}}_{\mathrm{B}}
                     + \hat{\mathcal{H}}_{\mathrm{int}}
\label{eq:Heff}
\end{equation}
where $\mu$ is the reduced mass of collision complex defined by
$\mu=m_{\mathrm{A}}m_{\mathrm{B}}/(m_{\mathrm{A}}+m_{\mathrm{B}})$,
$\hat{\mathcal{H}}_{\mathrm{A}}$ and $\hat{\mathcal{H}}_{\mathrm{B}}$ describe
separated A and B in an external magnetic field, and $\hat{\mathcal{H}}_{\mathrm{int}}$ describes the interaction between the collision partners. As mentioned in Sec. IIA the collision complex is described by the Jacobi vectors $\bm{R}$ and $\bm{r}$ in the BF frame. The embedding of the BF $z$ axis is chosen to coincide with the vector $\boldsymbol{R}$, and the BF $y$ axis is chosen to be perpendicular to the plane defined by the collision complex.

In  Eq. (\ref{eq:Heff}), $\hat{J}$ is the operator for the total angular momentum of the 
collision complex, $\hat{N}$ is the rotational angular momentum operator for molecule B, and $\hat{S}_{\mathrm{A}}$ and $\hat{S}_{\mathrm{B}}$
are the operators for the spin angular momenta of atom A and molecule B,
respectively. The orbital angular momentum operator of the collision complex in the BF frame is given by $\hat{l} = \hat{J}-\hat{N}-\hat{S}_{\mathrm{A}}-\hat{S}_{\mathrm{B}}$.
The separated atom Hamiltonian in the presence of an external magnetic field is given as $\hat{\mathcal{H}}_{\mathrm{A}}  = {\textrm{g}}_e\mu_\mathrm{B} \hat{S}_{{\rm
A},Z} B$, where $\textrm{g}_e$ is  the electron $\textrm{g}$-factor,
$\mu_\mathrm{B}$ is the Bohr magneton, $\hat{S}_{{\rm A},Z}$ gives the projection
of $\hat{S}_{\rm A}$ onto the magnetic field axis and $B$ is the magnitude of the external magnetic field. For a linear molecule such as SrOH($X^2\Sigma$), $\hat{\mathcal{H}}_{\mathrm{B}} = B_e
\hat{N}^2 + \gamma_\mathrm{SR} \hat{N} \cdot \hat{S}_{\rm B} + \textrm{g}_e\mu_\mathrm{B}
\hat{S}_{{\rm B},Z} B$, where $B_e$ is the rotational constant, and
$\gamma_\mathrm{SR}$ is the spin-rotation constant.
The last term in Eq. (\ref{eq:Heff}) describes the atom-molecule interaction,
including both the electrostatic interaction potential $\hat{V}$ and the magnetic dipole-dipole 
interaction $\hat{{V}}_\mathrm{dd}$ between the magnetic moments of
the atom and the molecule. The interaction potential $\hat{V}$ may be written
\begin{equation}
 \hat{V}(R,\theta) = \sum^{S_{\mathrm{A}}+S_{\mathrm{B}}}_{S=|S_{\mathrm{A}}-S_{\mathrm{B}}|}
  \sum^{S}_{\Sigma=-S} |S \Sigma \rangle \hat{V}^S(R,\theta) \langle S \Sigma|\,,%
 \label{eq:V}
\end{equation}
where total electronic spin $S$ is defined as $\hat{S} =\hat{S}_{\mathrm{A}}+\hat{S}_{\mathrm{B}}$. 
%\textcolor{red}{
%As described in Sec. IIA, we neglect the triplet contribution to the interaction potential.
%[Q). MASATO, DO WE ACTUALLY NEGLECT THE S = 1 TERM OR ASSUME THAT IT IS EQUAL TO THE S = 0 TERM? --------- A). Above my sentence is completely wrong. I revised as follows;] }
%\textcolor{blue}{
Our interest here is in collisions between rotationally ground-state SrOH molecules ($N=0$) with Li atoms initially in their maximally stretched Zeeman states $M_{S_\mathrm{A}}=M_{S_\mathrm{B}}=1/2$, where $M_{S_\mathrm{A}}$ and $M_{S_\mathrm{B}}$ are the projections of $\hat{S}_{\mathrm{A}}$ and $\hat{S}_{\mathrm{B}}$ onto the space-fixed $Z$-axis.
Following our previous work on Li-CaH \cite{pra11,Warehime:2015} we assume  that the non-adiabatic coupling between the triplet ($S = 1$) and the singlet ($S = 0$) Li-SrOH PESs can be neglected, and that the PESs are identical, {\it i.e.}  $\hat{V}^{S=0}(R,\theta)=\hat{V}^{S=1}(R,\theta)$.
%}
% \begin{equation}
%\hat{V}^S(R,\theta)
%          = \sum_ \lambda\hat{V}^S_\lambda(R) P_\lambda( \textup{cos}\: \theta).
% \label{eq:VS}
%\end{equation}
The dipolar interaction between the magnetic moments of the atom and molecule
may be written \cite{jcp12a}
\begin{equation}
 \hat{V}_\mathrm{dd}=
 -\textrm{g}_{e}^{2}\mu_{0}^{2}\sqrt{\frac{24\pi}{5}} \frac{\alpha ^2}{R^3}  \sum^{}_{q} (-)^qY_{2,-q}^*(\hat{\boldsymbol{R}}) [\hat{S}_{\rm A}\otimes \hat{S}_{\rm B}]_{q}^{(2)},
 %-\textrm{g}_{e}^{2}\mu_{0}^{2}\sqrt{6} \frac{\alpha ^2}{R^3} [\hat{s}_{\rm A}\otimes \hat{s}_{\rm B}]_{0}^{(2)},
 \label{eq:Hdip}
\end{equation}
%where $\mu_0$ is the magnetic permeability of free space and $\alpha$ is the
%fine-structure constant.
where   $\mu_0$ is the magnetic permeability of free space, $\alpha$ is the fine-structure constant and $ [\hat{S}_{\rm A}\otimes \hat{S}_{\rm B}]_{q}^{(2)}$ is the spherical tensor product of $\hat{S}_{\rm A}$ and $\hat{S}_{\rm B}$.

Following previous studies \cite{pra11,jcp10,jcp12a},  the total wave function of the Li-SrOH collision complex is expanded in a set of basis functions 
\begin{equation}\label{eq:totjbasis}
\left|JM\Omega\rangle|NK_N\rangle|S_{\mathrm{A}}\Sigma_{\mathrm{A}}\rangle|S_{\mathrm{B}}\Sigma_{\mathrm{B}}\right\rangle.
\end{equation}
Here, $\Omega$, $K_N$, $\Sigma_{\mathrm{A}}$ and $\Sigma_{\mathrm{B}}$ are the projections of $J$, $N$, $S_{\mathrm{A}}$ and $S_{\mathrm{B}}$ onto the BF quantization axis $z$, and  $\Omega=K_N+\Sigma_{\mathrm{A}}+\Sigma_{\mathrm{B}}$ is satisfied. Unlike $\Omega$, the projection of $J$ onto the space-fixed quantization axis $M$ is rigorously conserved for collisions in a DC magnetic field \cite{njp09,Krems_matrix}, so  the CC equations can be constructed and solved independently for each value of $M$. In Eq.~(\ref{eq:totjbasis})  $|JM\Omega\rangle=\sqrt{(2J+1)/8\pi^2}D^{J*}_{M\Omega}(\bar{\alpha},\bar{\beta},\bar{\gamma})$ is an eigenfunction of the symmetric top, and the Wigner $\mathit{D}$-functions $D^{J*}_{M\Omega}(\bar{\alpha},\bar{\beta},\bar{\gamma})$ depend on the Euler angles $\bar{\alpha}$, $\bar{\beta}$ and $\bar{\gamma}$, which specify the position of the BF axes $x$, $y$ and $z$ in the SF frame. The rotational degrees of freedom of SrOH in the BF frame are described by the functions $|NK_N\rangle$, which can be expressed using the spherical harmonics as $\sqrt{2\pi} Y_{NK_N}(\theta,0)$. The matrix elements of the effective Hamiltonian  in the total angular momentum  basis (\ref{eq:totjbasis}) are evaluated  as described elsewhere \cite{jcp10}.

\begin{table}[t]
\caption{Spectroscopic constants of SrOH (in cm$^{-1}$) and masses of the collision partners (in~a.m.u.) used in scattering calculations.}
\begin{tabular} {p{5em} cc}  \hline \hline
Parameter                 & Value            \\  \hline  \hline
 $B_e$                      &   0.24633     \\
 $\gamma_{\mathrm{SR}}$               &   $2.4275\times 10^{-3}$  \\
 $m_\mathrm{SrOH}$   & 104.9083586           \\
 $m_\mathrm{Li}$        &   7.01600455           \\  \hline
\end{tabular}
\label{tab:parameters}
\end{table}

The molecular parameters used in scattering calculations are listed in Table \ref{tab:parameters}. The size of the basis set is governed by the truncation parameters of $J_\mathrm{max}$ and $N_\mathrm{max}$ which show the maximum quantum numbers of the  total angular momentum $J$ of the collision complex Li-SrOH and the rotational angular momentum $N$  of SrOH in the basis set. Unless stated otherwise, all scattering calculations are carried out with $J_\mathrm{max}=3$. The convergence properties with respect to $J_\mathrm{max}$ are examined in the Appendix. Due to the strong anisotropy of the Li-SrOH interaction, a large number of rotational channels must be included in the basis set to obtain converged results. Furthermore, the rotational constant  of SrOH is $\sim$20 times smaller than that of CaH, which results in  larger values of $N_\mathrm{max}$ for Li-SrOH compared to  Li-CaH ($N_\mathrm{max}=55$ \cite{pra11}). Indeed, we found that using $N_\mathrm{max}=115$ is necessary to obtain the cross sections converged to within 2\% (see the Appendix). 

The numerical procedures used in this work are essentially the same as those employed in our previous study of Li-CaH  collisions \cite{pra11} as explained in detail elsewhere \cite{jcp10,jcp12a}. In brief, 
the CC equations are  solved numerically using the log-derivative propagator method  \cite{prop_1, prop_2} 
on an equidistant radial grid from 
%$R=4.0$ $a_0$ to $R_1$ with $R_1=9.5$ $a_0$ for $B > 10$~G and $R_1=22.7$ $a_0$ for $B < 10$~G using a step size of $0.00189$ $a_0$.  Airy propagation is employed for $R_1 <R <R_\text{max}$ with $R_\text{max}=280.0$ $a_0$ for $B > 10$~G and $R_\text{max}=1322.8$~$a_0$ for $B < 10$~G.  
$R_\text{min}=4.0$ $a_0$ to $R_\text{mid}$ with $R_\text{mid}=9.5$ $a_0$ for $B > 10$~G and $R_\text{mid}=22.7$ $a_0$ for $B \le 10$~G using a step size of $0.00189$ $a_0$. 
Airy propagation is employed for $R_\text{mid} \le R \le R_\text{max}$ with $R_\text{max}=280.0$ $a_0$ for $B > 10$~G and $R_\text{max}=1322.8$~$a_0$ for $B \le 10$~G.  
Propagating the log-derivative matrix out to very large values of $R_\text{max}$ is necessary to  maintain the numerical accuracy of quantum scattering calculations on systems with long-range anisotropic interactions at low magnetic fields  \cite{JanssenNH}. 

After  propagating  the log-derivative matrix out to a sufficiently large $R=R_\mathrm{max}$ where the interaction potential becomes negligible, the  matrix is transformed from the total angular momentum representation to a basis set in which $\hat{\mathcal{H}}_{\mathrm{A}}$, $\hat{\mathcal{H}}_{\mathrm{B}}$ and $\hat l^2$ are diagonal. The resultant log-derivative matrix is matched to the scattering boundary conditions to obtain the $S$-matrix, and the elastic and inelastic cross sections are extracted from the $S$-matrix as described in Ref. \cite{jcp10}.  

%Given the high density of rovibrational states of Li-SrOH and a pronounced anisotropy of the Li-SrOH PES (see Sec. A above), going beyond the rigid-rotor approximation would be a formidable task. However, we expect that the inclusion of the bending and a few lowest excited vibrational modes in the basis set should be feasible.  

%Within the rigid-rotor approximation, the Hamiltonian of the Li-SrOH collision complex becomes the same as those employed previously  for Li-CaH \cite{LiCaH}, with the important difference that the  The Hamiltonian takes the form

%theory and computational methodology of
%However, the rotational constant of SrOH is about 10 times smaller than that of CaH, which makes obtaining converged results more challenging that for Li-CaH (see below for convergence testst).

\subsection{Quantum capture calculations}

The Li($^2$S) + SrOH($^2\Sigma$) $\to$ LiOH + Sr chemical reaction can proceed through the singlet and triplet pathways. The singlet pathway gives the ground-state products and is exothermic by $\sim$0.32~eV~\cite{JANAF}. The triplet pathway involves the  $^3A$ PES (the entrance channel of which is calculated in Sec. IIA) and correlates to the Sr atom excited to the $^3$P state, which lies 1.8 eV above the ground state. Thus, the chemical reaction of spin-polarized reactants is energetically forbidden at low collision energies. However, spin-nonconserving interactions  (such as spin-orbit  and hyperfine) may induce non-adiabatic transitions at short range \cite{Warehime:2015}, which are not accounted for within our reduced-dimensional single-state model. An upper bound to the rate of these transitions is given by the capture rate, {\it i.e.}, the rate of reactant penetration to the short-range region as defined classically by the Langevin model. To estimate this rate, we applied here a quantum version of the statistical adiabatic channel model~\cite{VMU} implemented as described in Ref.~\cite{NJP}. 

In brief, we use a simplified atom-molecule Hamiltonian (\ref{eq:Heff}) without the spin-rotation coupling and Zeeman interactions. The adiabatic channel potentials are obtained by diagonalizing the Hamiltonian, at fixed atom-molecule separations $R$, in the symmetry-adapted rigid rotor function basis set. Since we are only interested in the channels correlating to the ground rotational state of the SrOH reactant, the single lowest-energy root was retained for each total angular momentum quantum number $J$ ($l \equiv J$ in this case). We found that 50 basis functions with $N_\mathrm{max}=49$ give  results converged to within 2\% for the desirable $N=0$ adiabatic channel near the bottom of the potential well at $R = 5.3$ $a_0$. 

%{\tt I would not plot adiabatic channels: nothing spectacular?}

To calculate the quantum capture probabilities  for $J \le 20$, we use the modified Truhlar-Kupperman finite difference method~\cite{TK} as described in Refs.~\cite{VMU,NJP} on a grid of collision energies extending from 10$^{-11}$ to 1000 cm$^{-1}$. Inner capture boundary conditions are applied at  $R=R_0$ within the short-range region. We used 6 values of $R_0 \in [5.7,13.2]a_0$ to obtain the average capture probability at each collision energy. The classical capture probabilities are determined from the height of the centrifugal barrier for each $J \le 40$ \cite{NJP}.

\section{Results}

\subsection{Elastic and inelastic cross sections}

\begin{figure}[ht]
\begin{center}
\includegraphics[height=0.33\textheight,keepaspectratio]{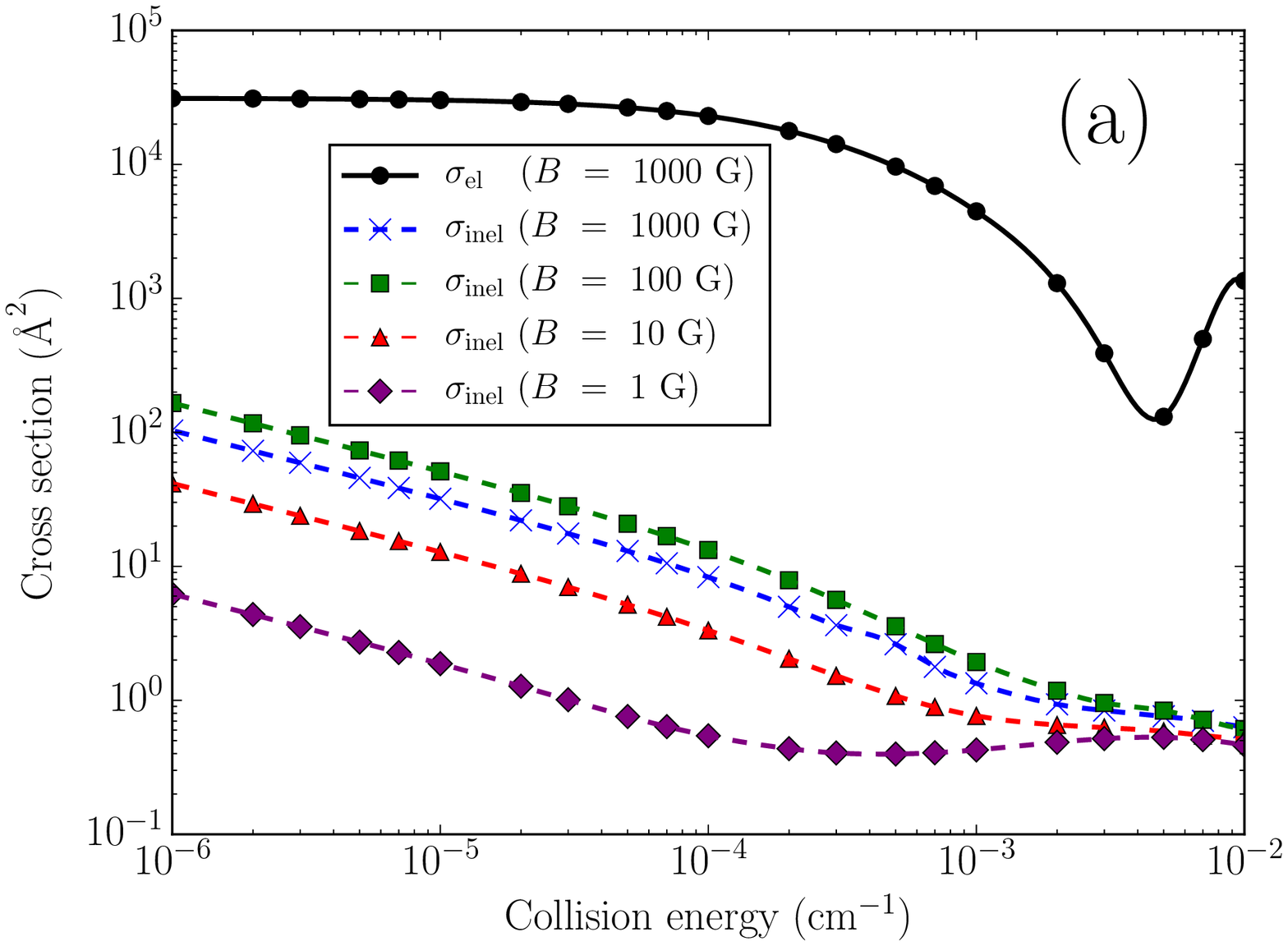}
\includegraphics[height=0.33\textheight,keepaspectratio]{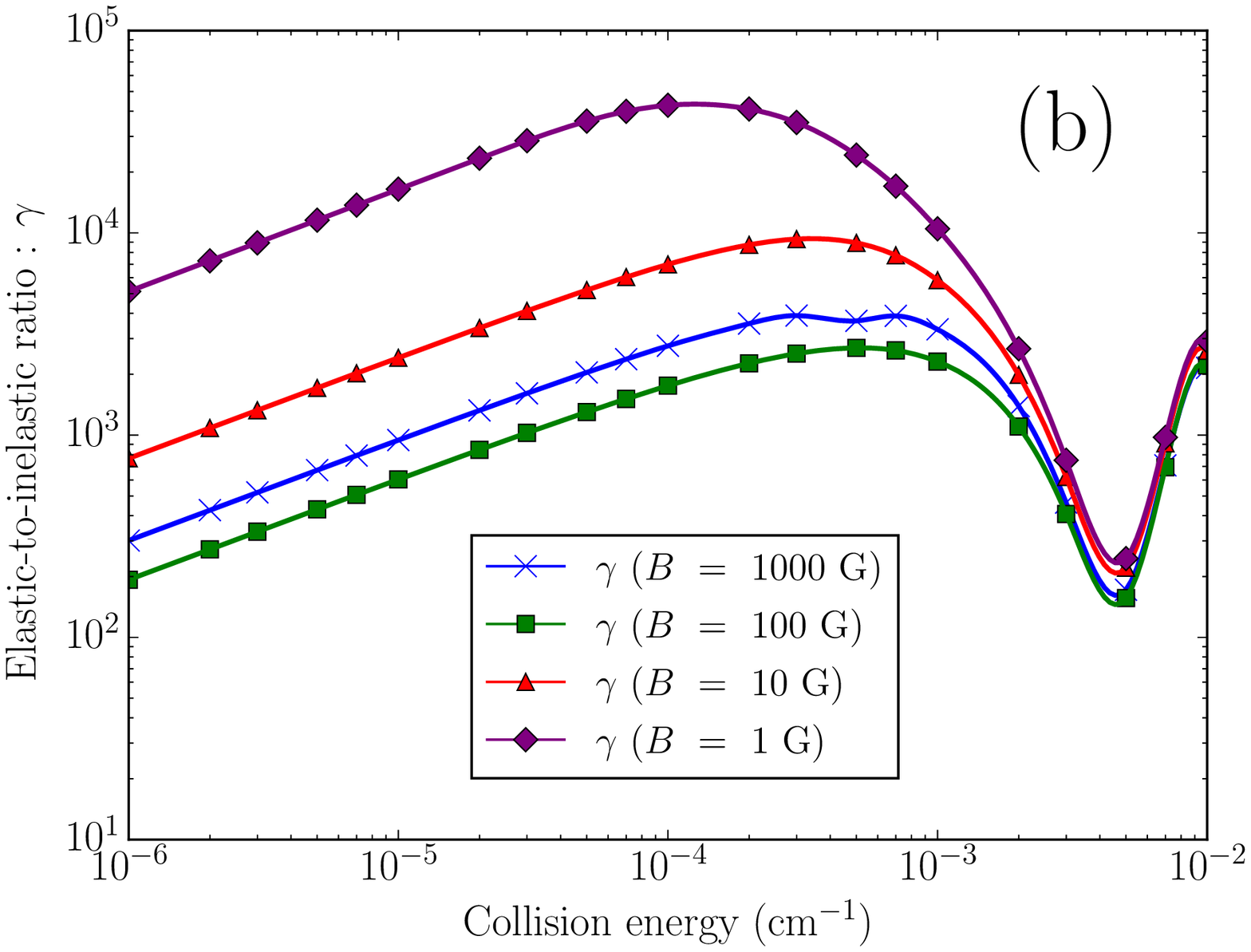}
\end{center}
\caption{ (a) Cross sections for elastic scattering and inelastic spin relaxation in spin-polarized Li~+~SrOH collisions plotted as functions of collision energy for the external magnetic field of 1 G (diamonds), 10 G (triangles), 100 G (squares), 1000 G (crosses). The elastic cross section displays a very weak magnetic field dependence. (b) The ratios of elastic and inelastic cross sections as functions of collision energy for the same values of the magnetic field as in (a). }
\label{figure2}
\end{figure}

Figure\ \ref{figure2} (a) shows the elastic ($\sigma_\mathrm{el}$) and inelastic ($\sigma_\mathrm{inel}$) cross sections for fully spin-polarized Li~+~SrOH collisions as functions of collision energy for the external magnetic fields of 1 G, 10 G, 100 G and 1000 G. Due to the very weak magnetic field dependence of  the elastic cross section, only  the $B=1000$~G result  is shown in the figure. The inelastic cross section increases significantly as the magnetic field increases from 1~G to 100~G, especially in the ultracold $s$-wave regime (the field dependence will be explored later in this section).
The inelastic cross sections as a function of collision energy remain smooth and small over the entire energy range considered. As mentioned in the Introduction, for sympathetic cooling to be effective, the ratio of elastic to inelastic cross sections $\gamma=\sigma_\mathrm{el}/\sigma_\mathrm{inel}$ must exceed 100. Figure\ \ref{figure2}(b) shows that the calculated values of $\gamma$ do exceed 100  throughout the whole energy range except in the vicinity of $E_C=5.0 \times 10^{-3}$ cm$^{-1}$.
%  suggesting good prospects for sympathetic cooling of SrOH molecules with magnetically co-trapped Li atoms.

\begin{figure}[ht]
\begin{center}
\includegraphics[height=0.35\textheight,keepaspectratio]{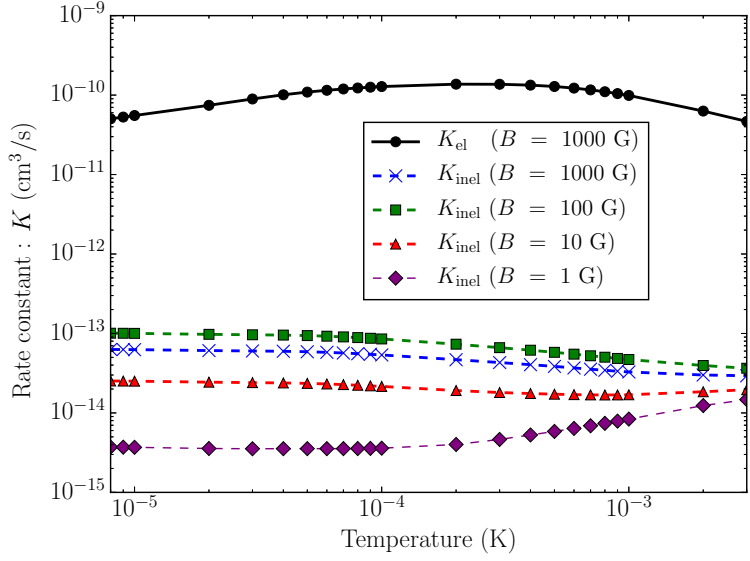}
\end{center}
\caption{Temperature dependence of the rate constants for elastic scattering (circles)  and inelastic spin relaxation in spin--polarized  Li~+~SrOH collisions calculated for the magnetic field values of 1 G (diamonds), 10 G (triangles), 100 G (squares), 1000 G (crosses).}
\label{figure3}
\end{figure}

Figure\ \ref{figure3} shows the temperature dependence of the rate constants for elastic scattering and spin relaxation. The rate constant is an energy averaged property obtained by the convolution of the cross sections with the Maxwell-Boltzmann distribution function. As such, the behavior of the rate constant as function of temperature tends to be monotonous. Importantly, collision-induced spin relaxation in Li~+~SrOH collisions occurs more than two orders of magnitude slower than elastic scattering, suggesting favorable prospects for sympathetic cooling of SrOH molecules with Li atoms in a magnetic trap.

As shown in Fig.\ \ref{figure2}(a), the inelastic cross section decreases dramatically as the magnetic field is reduced from 100~G to 1~G in the ultracold $s$-wave regime. The suppression of spin relaxation  is a consequence of conservation of parity and the total angular momentum projection $M$ \cite{VolpiBohn}, which dictate that  inelastic spin relaxation of the incoming $s$-wave channel must be accompanied by a change of the orbital angular momentum from $l=0$  to $l=2$. 
If the energy difference between the initial and final channels is small enough due to the small Zeeman splitting in a weak magnetic field, the height of the $d$-wave centrifugal barrier in the final channel can be larger than the initial kinetic energy in the incoming channel. 
Under such conditions,  spin relaxation occurs by tunnelling under the $d$-wave centrifugal barrier, and is strongly suppressed.
We note that, as discussed in Sec.~IIB below, this mechanism only applies to indirect spin relaxation induced by the intramolecular spin-rotation interaction.

\begin{figure}[ht]
\begin{center}
\includegraphics[height=0.35\textheight,keepaspectratio]{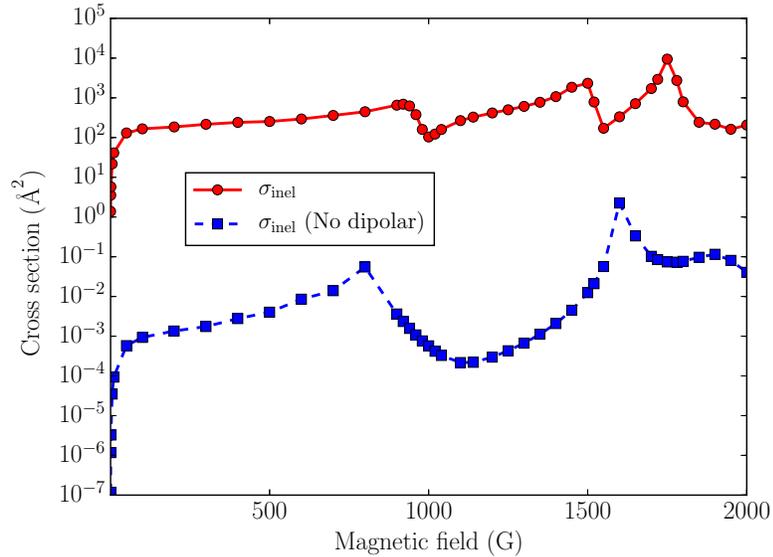}
\end{center}
\caption{Magnetic field dependence of the inelastic cross sections for Li~+~SrOH calculated with (full line with circles) and without (dashed line with squares) the magnetic dipole-dipole interaction. The collision energy of $1.0 \times 10^{-6}$ cm$^{-1}$.}
\label{figure4}
\end{figure}

In Fig.\ \ref{figure4}, we plot the magnetic field dependence of the inelastic cross sections calculated for the collision energy of $10^{-6}$ cm$^{-1}$. We observe two broad asymmetric resonance profiles, around which  
 the inelastic cross sections are reduced dramatically. This suggests the possibility of suppressing spin relaxation in Li~+~SrOH collisions by tuning the DC magnetic field as noted previously for He-$\mathrm{O_2}$ \cite{JMH_inel}. We note that, despite the high density of rovibrational states of the Li-SrOH collision complex, only a few resonances are observed in the inelastic cross section below 2000 G. This suggests that most of the states of the complex are decoupled from the incident spin-polarized collision channel. A similar magnetic field dependence is observed in ultracold collisions of spin-polarized alkali-metal atoms \cite{Burke98,BurkeThesis} and O$_2$($^3\Sigma$) molecules \cite{TscherbulNJP09,O2O2collisions2011}, which display a lower resonance density in non-spin-polarized initial channels. 

%\textcolor{blue}{ (Now, I think this paragraphs is not necessary for this paper and may cause confusion for readers. Please remove if you also think so. )}
%To quantitatively explain the origin of the observed resonance density, it is necessary to solve the Schr\"{o}dinger equation for (quasi) bound states of the collision complex to count the resonances, including extremely narrow resonances and overlapping resonances. Although such calculations are beyond the scope of this work, we note that such narrow resonances  would be different from the  wide resonances shown in Fig.\ \ref{figure4}. The presence of the narrow resonances would indicate that the resonance states are not coupled strongly with each other, but could be classified into (hierarchical) groups distinguished by their characteristics such as the wave functions and widths, similarly to the case of Li-CaH and Li-CaF \cite{JMH_nonChaos}.

%\textcolor{red}{ [If I can successfully complete the converged calculation using larger $R_\mathrm{max}$ for weak field (around and less than 1G), it might be possible to discuss the threshold behavior of inelastic cross sections as a function of $B$, and compare the results  of NH-NH by Groenenboom et al.] }

\subsection{Direct vs. indirect spin relaxation mechanisms}

In general, inelastic spin relaxation in cold collisions of $^2\Sigma$ molecules in their ground rotational states with $^2\mathrm{S}$ atoms is mediated by two mechanisms, direct and indirect. The direct mechanism is due to the long-range intermolecular magnetic dipole-dipole interaction $\hat{{V}}_\mathrm{dd}$ given by  Eq.~(\ref{eq:Hdip}). The indirect mechanism is due to a combined effect of the intramolecular spin-rotation interaction and the coupling between the rotational states of the molecule induced by the anisotropy of the interaction potential   \cite{Krems_spinrot}. As shown in Fig.~1, the anisotropy of the interaction potential is strong in the range of small atom-molecule distances $R$; thus the indirect mechanism operates at short range.  

In order to compare these mechanisms, we plot in Figs.\ \ref{figure4} and 5(a) the inelastic cross sections calculated with and without the magnetic dipole-dipole interaction. Omitting the magnetic dipole-dipole interaction leads to a dramatic reduction of the inelastic cross section over the entire magnetic field range (including near scattering resonances), which strongly suggests that spin relaxation in spin-polarized Li~+~SrOH collisions is driven by the direct mechanism.
As shown in Fig. 5(a),  the  indirect mechanism becomes more efficient with increasing collision energy; however, the direct mechanism   dominates even at  the highest collision energy considered.

It is instructive to  compare the efficiency of  the  indirect spin relaxation mechanism  in collisions of light (CaH) and heavy  (SrOH) molecular radicals with Li atoms.   The  inelastic cross sections calculated in the absence of the magnetic dipole-dipole interaction are similar in magnitude  ($5.7\times 10^{-4}$ \AA~ for Li~+~SrOH and $10^{-3}$ \AA~for Li~+~CaH \cite{pra11} at $E_C=10^{-6}$ cm$^{-1}$ and $B=0.1$~T).  At first glance, this is surprising because $\sim$120 excited rotational states contribute to the indirect spin relaxation mechanism for Li~+~SrOH, as opposed to only $\sim$50 rotational states for Li~+~CaH. As a result, the number of third (and higher)-order contributions to the  Li~+~SrOH inelastic scattering amplitude is expected to be significantly larger than for Li~+~CaH, leading one to expect the indirect spin relaxation mechanism to be more efficient for Li~+~SrOH. However, the  spin-rotation constant of SrOH is 10 times smaller than that of CaH, so each contribution to the Li~+~SrOH scattering amplitude is suppressed by a factor of 10.  This suppression compensates for the larger number of contributing terms  for Li~+~SrOH, providing a qualitative explanation for the comparable efficiency of indirect spin relaxation mechanisms in collisions of light and heavy  molecular radicals.

% which  (more than twice as many as needed for Li-CaH).  and hence one should expect more excited rotational states should be involved in Li-SrOH collisions, causing more efficient transfer of spin population via excited rotational states.

 Figures 5(b) and (c) compare the  incoming partial wave contributions to the inelastic cross sections  calculated with and without  the  magnetic dipole-dipole interaction. For the indirect spin relaxation mechanism, the incoming $p$-wave contribution tends to exceed the incoming $s$-wave contribution in both the $s$-wave and multiple-partial-wave regimes as discussed in Refs.~\cite{VolpiBohn, Krems_matrix}. 
In contrast, the cross sections  calculated  with the magnetic dipole-dipole interaction included display a more  conventional partial wave structure, with the incoming $s$-wave contributions being  dominant below $E_C=10^{-3}$ cm$^{-1}$ and all  incoming partial wave components becoming comparable  at higher collision energies.  
This explains the diminishing role of the indirect spin relaxation mechanism with decreasing collision energy evident in Fig.~5(a).

\subsection{Quantum capture rates}

Figure~\ref{fig:capture} shows the Li~+~SrOH capture rate constant as a function of temperature, with quantum and classical results shown by lines and symbols, respectively. At $T \to 0$, the quantum rate obeys the Wigner threshold law for $s$-wave scattering. The crossover to multiple scattering regime, which occurs at {\it ca.} 100 $\mu$K, manifests itself as a shallow minimum in the temperature dependence of the total capture rate. The total classical capture rate exhibits the expected divergence as $T \to 0$ due to the lack of quantum reflection in the barrierless $s$-wave scattering channel. On the other hand, neglecting tunnelling leads to a faster decline of the contributions from higher partial waves as the temperature decreases. A combination of these two effects makes the classical capture approximation quite reasonable down to the temperatures as low as 300 $\mu$K. Overall, the magnitude of the Li~+~SrOH  reaction rate and its temperature dependence  are very similar to those calculated previously for the Li + CaH $\to$ LiH + Ca chemical reaction~\cite{NJP}. However, the larger reduced mass of the Li-SrOH collision complex and its stronger long-range dispersion forces make the crossover effect more pronounced and the classical approach more reliable for the $J > 0$ partial waves.

We note that our calculated capture rates cannot be directly compared to those obtained from  quantum scattering calculations using the standard boundary conditions. However, the large magnitude of the capture rates indicates that the  Li + SrOH $\to$ LiOH~+~Sr chemical reaction should be the dominant loss channel for the reactants colliding in non fully spin-polarized initial states. Assuming that the long-range behavior of the singlet and triplet PESs is identical, we obtain an upper bound to the reaction rate as 3 $\times$ 10$^{-10}$ cm$^3$/s, one third of the value shown in Fig.~\ref{fig:capture}, and 3-4 orders of magnitude larger than the spin relaxation rate for fully spin-polarized Li~+~SrOH collisions shown in Fig. 3. Thus, spin polarization of the reactants can be considered as a way to suppress  inelastic and reactive losses in cold Li~+~SrOH collisions.

\section{Summary and outlook}

We have studied the collisional properties of  ultracold spin-polarized mixtures of SrOH molecules with Li atoms using reduced-dimensional quantum scattering calculations and a newly developed, highly anisotropic triplet PES of the Li-SrOH collision complex. We present the elastic and inelastic collision cross sections over a wide range of collision energies $(10^{-6}$-1~K) and magnetic fields (1-1000~G) along with the quantum and classical capture rates, which give an upper limit to the total Li+SrOH reaction rate. We find that inelastic spin relaxation in fully spin-polarized Li~+~SrOH collisions is strongly suppressed (with the ratio of elastic to inelastic collision  rates $\gamma>10^2$-$10^3$), suggesting good prospects for sympathetic cooling of spin-polarized SrOH molecules with Li atoms in a magnetic trap. In the context of rapid experimental progress in buffer-gas cooling and Sisyphus laser cooling   of polyatomic radicals  \cite{Ivan1,Ivan2,Ivan3}, our results open up the possibility of sympathetic cooling of polyatomic molecules with magnetically co-trapped ultracold alkali-metal atoms, potentially leading to new advances  in low-temperature chemical dynamics and spectroscopy of large molecules  in the gas phase \cite{prl11,IvanCPC}.

In future work, we intend to explore the sensitivity of scattering observables to small uncertainties of the Li-SrOH interaction PES  (preliminary calculations indicate that the main conclusions of this work are robust against the uncertainties).
 It would also be instructive to study  the effect of the SrOH vibrational modes and singlet-triplet interactions neglected here \cite{Warehime:2015} on cold collisions of SrOH molecules with alkali-metal atoms in arbitrary initial quantum states. Such interactions could be particularly important for heavier coolant atoms, such as K and Rb, whose use in sympathetic cooling experiments may be preferable for experimental reasons.

 \section*{Acknowledgements}
 We are grateful to John Doyle and Ivan Kozyryev for stimulating discussions. This work was supported by the NSF (grant No. PHY-1607610).
 
 % which could be more advantageous to use as coolant atoms from an experimental viewpoint.

\appendix
\section{Basis set convergence of scattering observables}

Here, we explore the convergence of Li~+~SrOH scattering cross sections with respect to the truncation parameters   $N_\text{max}$ and $J_\text{max}$. First, we check the convergence with respect to the maximum rotational state included in the basis set $N_\text{max}$. As pointed out in Sec.~IIB, the small rotational constant of SrOH along with the large well depth and strong anisotropy of the Li-SrOH interaction lead to  a large value of $N_\text{max}$ required for  convergence. Figure~\ref{fig:figureA1} shows the cross sections as a function of $N_\text{max}$ at the collision energy of $1.0 \times 10^{-6}$~cm$^{-1}$  and the magnetic field of 100 G with  $J_\text{max}=1$. We observe rapid oscillations in the calculated  cross sections  until  $N_\text{max}\sim 95$. Even after the oscillations cease at  $N_\text{max}>100$, monotonous but non-negligible change of the cross sections continues until  $N_\text{max}=105$. We note that there seems to be no correlation between the behavior of the elastic and inelastic cross sections as a function of  $N_\text{max}$ in the  region of strong oscillations ($60<N_\text{max}<95$). The convergence patterns observed at higher collision energies ({\it e.g.} $10^{-3}$~cm$^{-1}$)  resemble those shown in Fig.~\ref{fig:figureA1}, with the oscillations becoming  less pronounced.   %Also we do not observe any special correlation between Figures \ref{fig:figureA1} and \ref{fig:figureA2} in such region. 
%Thereby, it is difficult to estimate in advance the impact of using non-converged basis set onto the resultant values of the observables including the elastic-to-inelastic ratio. 
We find that using $N_\text{max}=115$ gives both the elastic and inelastic cross sections converged to within 2\%.
%, hence, $N_\text{max}=115$ is used for all calculations reported herein.  

To test the convergence of scattering observables with respect to the maximum value of the total angular momentum in the basis set $J_\text{max}$, we plot in Fig. \ref{fig:figureA2} the elastic and inelastic cross sections as a function of collision energy calculated for $J_\text{max}=3$ and 4.
Note that since the couplings between the adjacent $J$-blocks become stronger with increasing the $B$-field \cite{jcp10},  using $B=1000$ G provides a more stringent convergence test than using $B=100$~G.
 As the computational cost of the $J_\text{max}=4$ calculations is very large, we limit the calculations  to 7 representative collision energies spanning the range $10^{-6}-10^{-2}$~K. Figure \ref{fig:figureA2} shows that adequate convergence of the inelastic cross sections is achieved with $J_\text{max}=3$ at all collision energies. The observed convergence for $J_\text{max}=3$ implies  the smallness of the incoming $f$-wave contributions to the  inelastic cross sections (described by adding the $J=4$ block in the basis set).
It also implies that the couplings between the incoming $f$-wave and $p$-wave scattering states in the entrance and exit collision channels are not critically important.

%\textcolor{red}{ }

\clearpage

\begin{figure}[ht]
\begin{center}
\includegraphics[height=0.27\textheight,keepaspectratio]{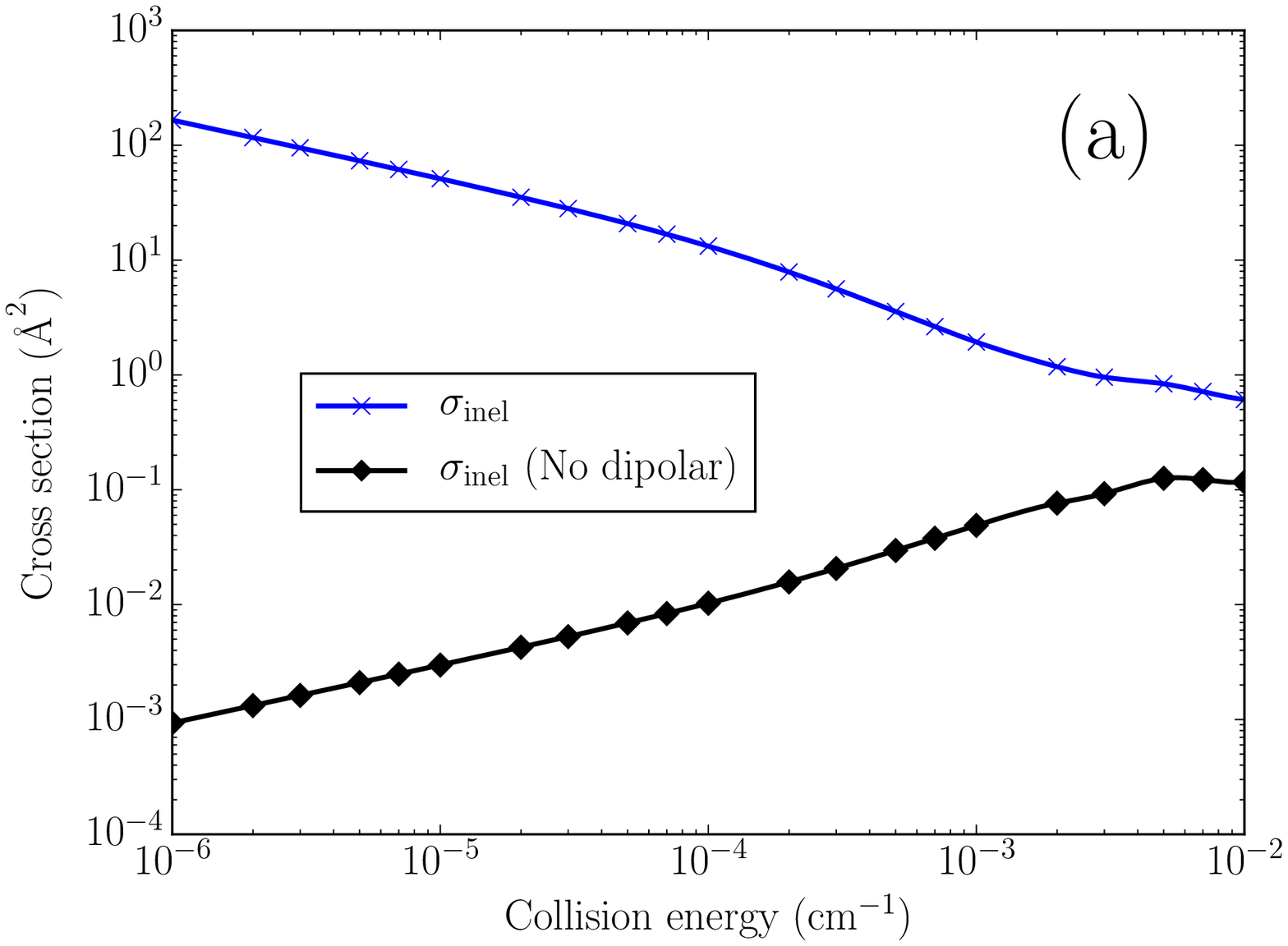}
\includegraphics[height=0.27\textheight,keepaspectratio]{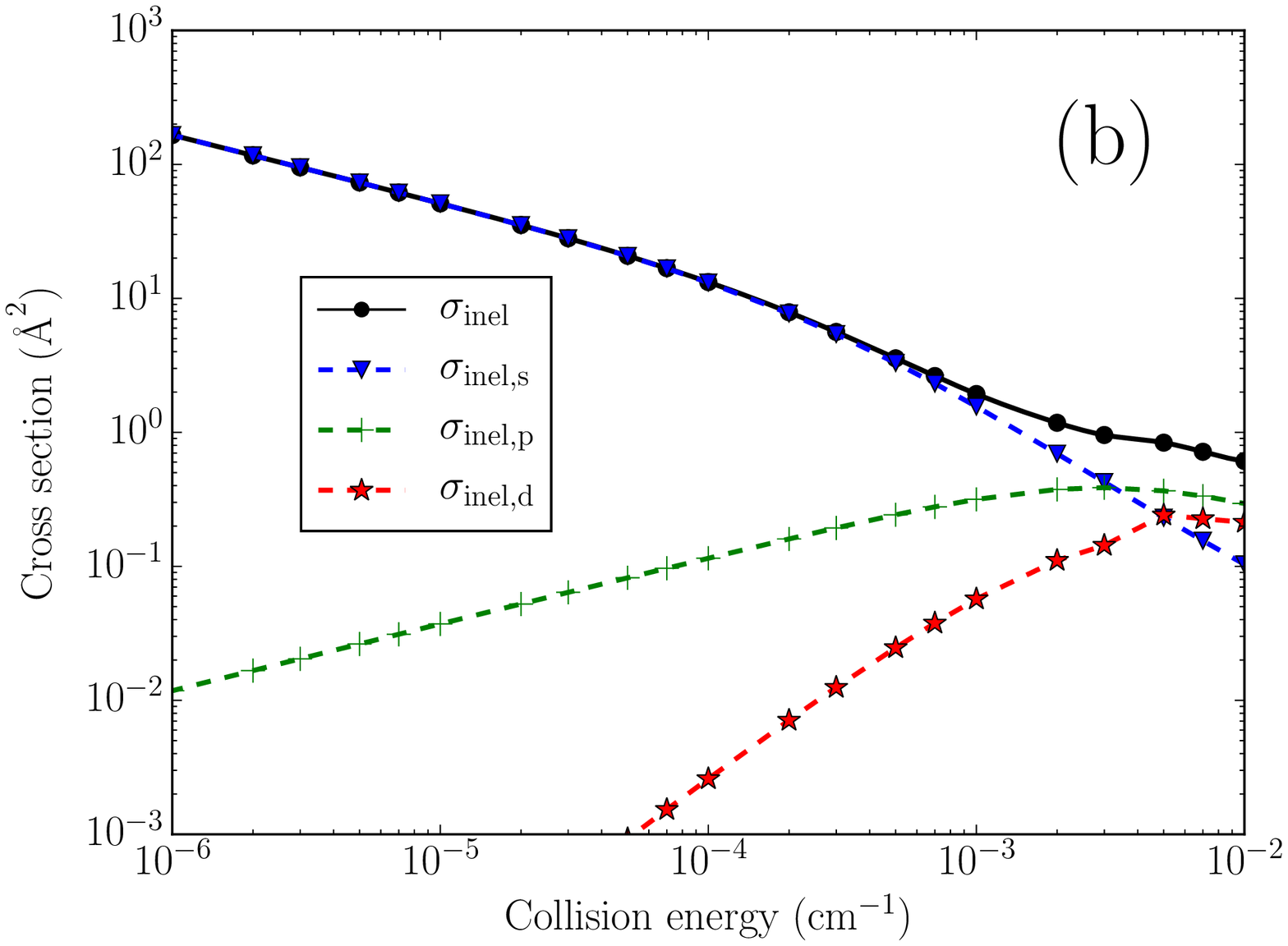}
\includegraphics[height=0.27\textheight,keepaspectratio]{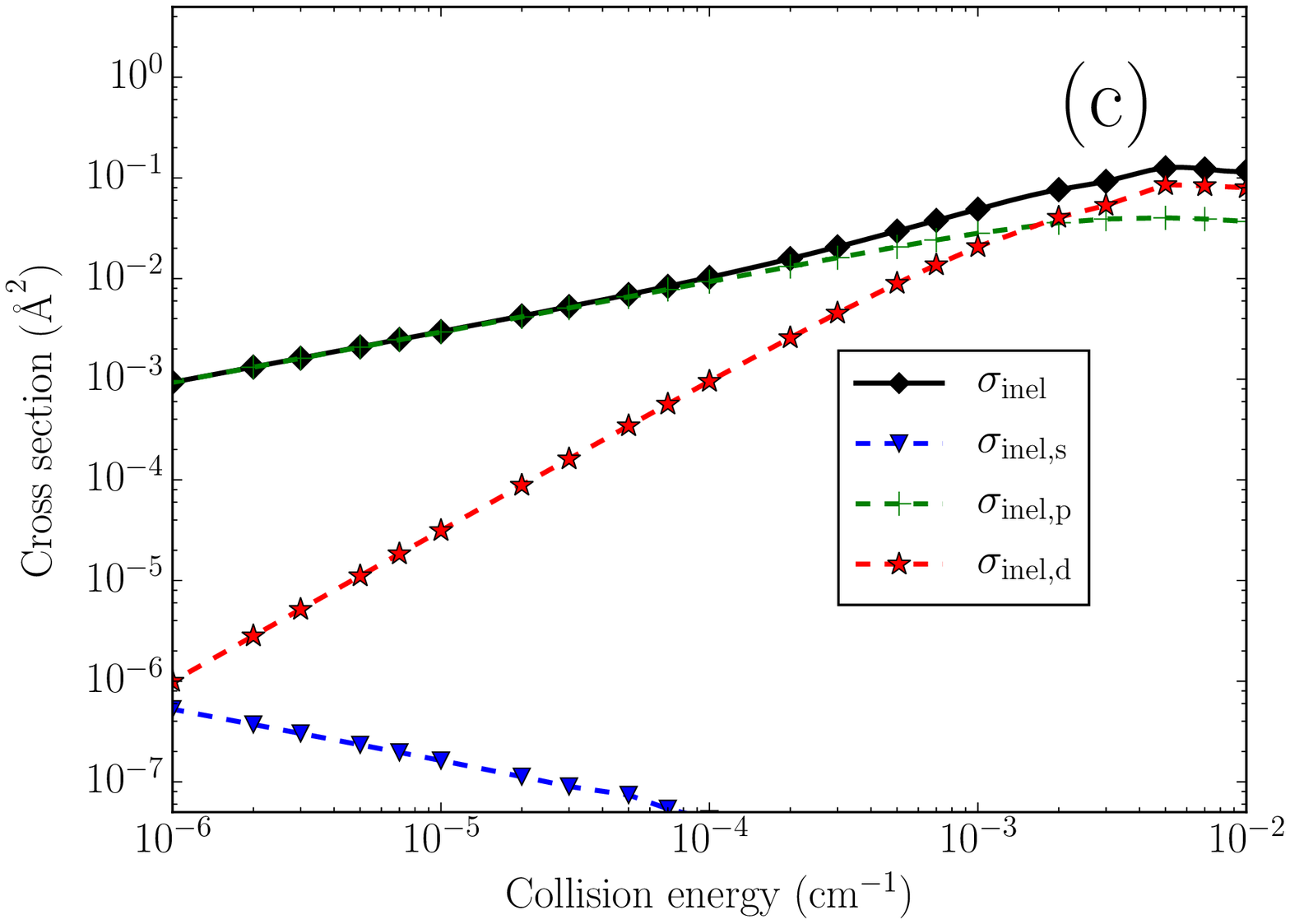}
\end{center}
\caption{(a) Collision energy dependence of the inelastic cross section calculated with (crosses) and without (diamonds) the magnetic dipole-dipole interaction for the magnetic field of 100 G. (b) Incoming partial wave decomposition of the inelastic cross section. (c) Same as in panel (b) but calculated without the magnetic dipole-dipole interaction.}
\label{figure5}
\end{figure}

\begin{figure}[ht]
\begin{center}
\includegraphics[height=0.40\textheight,keepaspectratio]{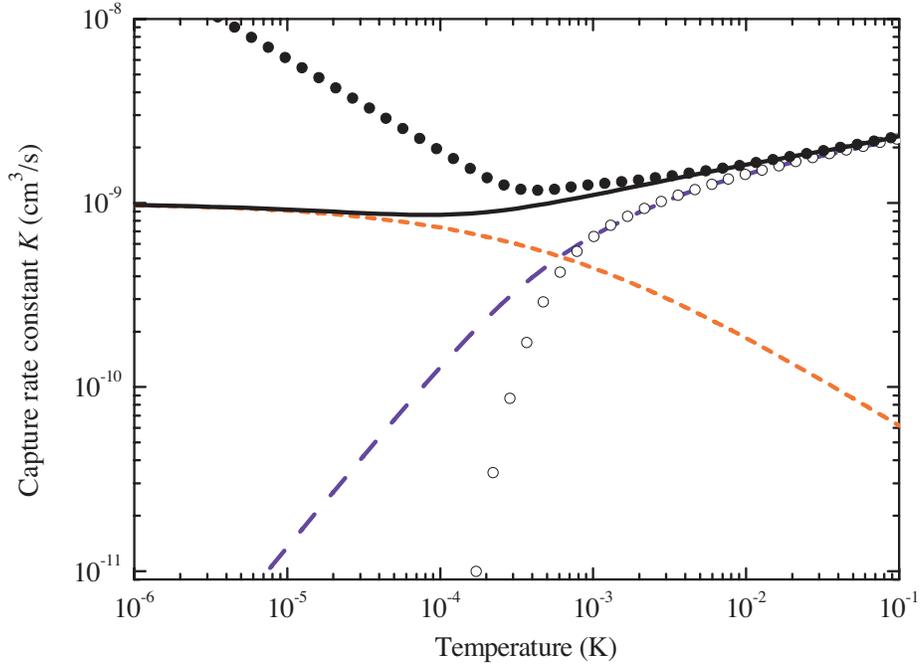}
\end{center}
\caption{Adiabatic capture rate constant for the Li~+~SrOH chemical reaction calculated as a function of collision energy in the absence of an external magnetic field. Quantum calculations: total rate constant (black solid line), $s$-wave rate constant (red dotted line) and the higher partial wave contribution (blue dashed line). Classical calculations: total rate constant (dots) and the $J > 0$ contribution (open circles). }
\label{fig:capture}
\end{figure}

%\setcounter{figure}{0}

%%%%%%%%%%%%%%%%%%%%%%%%%%%%%%%%%%%%%%%%%%%%%%%%%%%%%%%%%%%%%%%%%%%%%%%%%%%%%%%%%%%%%%%%%%%%%%%
%%%%%%%   PLEASE COPY and PASTE following section %%%%%%%%%%%%%%%%%%%%%%%%%%%%%%%%%%%%%%%%%%%%%%%%

\begin{figure}[ht]
\begin{center}
\includegraphics[height=0.35\textheight,keepaspectratio]{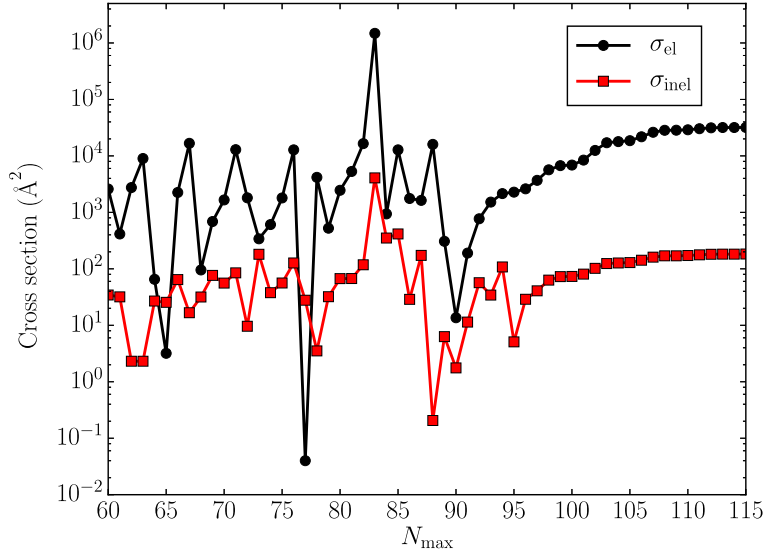}
\end{center}
\caption{Convergence of the elastic and inelastic cross sections for spin-polarized Li~+~SrOH collisions with respect to the number of rotational states included in the basis set at the collision energy of $1.0 \times 10^{-6}$ cm$^{-1}$. The  magnetic field is 100 G and $J_\text{max}=1$. }
\label{fig:figureA1}
\end{figure}

%\begin{figure}[ht]
%\begin{center}
%\includegraphics[height=0.35\textheight,keepaspectratio]{FigA2.eps}
%\end{center}
%\caption{Convergence of the elastic and inelastic cross sections for Li-SrOH with respect to the number of rotational states included in the basis set at the collision energy of $1.0 \times 10^{-3}$ cm$^{-1}$. The  magnetic field is 100 G and $J_\text{max}=1$. }
%\label{fig:figureA2}
%\end{figure}

\begin{figure}[ht]
\begin{center}
\includegraphics[height=0.35\textheight,keepaspectratio]{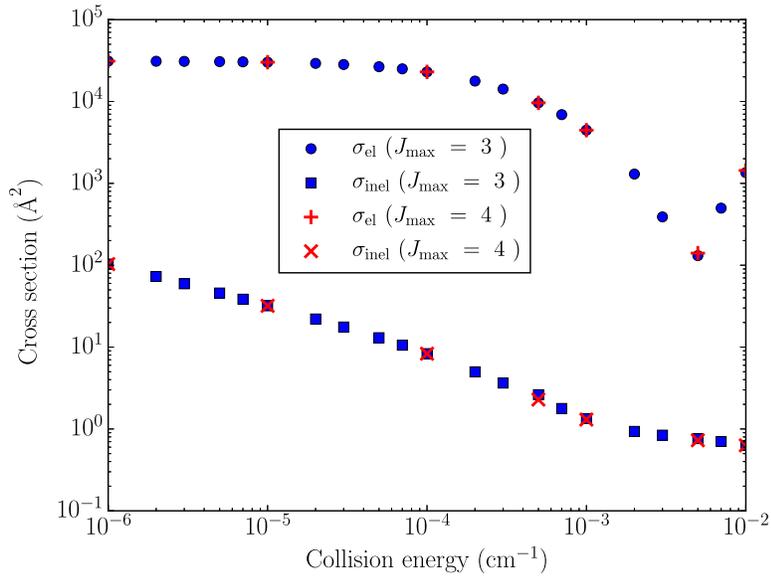}
\end{center}
\caption{Convergence of the elastic and inelastic cross sections for spin-polarized Li~+~SrOH collisions with respect to the number of total angular momenta included in the basis set: $J_\text{max}=3$ (circles and squares) and $J_\text{max}=4$ (pluses and crosses). The  magnetic field is 1000 G. }
\label{fig:figureA2}
\end{figure}

%%%%%%%%%%%%%%%%%%%%%%%%%%%%%%%%%%%%%%%%%%%%%%%%%%%%%%%%%%%%%%%%%%%%%%%%%%%%%%%%%%%%%%%%%%%%%%%
%%%%%%%%%%%%%%%%%%%%%%%%%%%%%%%%%%%%%%%%%%%%%%%%%%%%%%%%%%%%%%%%%%%%%%%%%%%%%%%%%%%%%%%%%%%%%%%

\end{document}